\documentclass[]{interact}

\usepackage{epstopdf}
\usepackage{graphicx}

\usepackage{dsfont}
\usepackage{xcolor}
\usepackage{tcolorbox}
\usepackage{cancel}
\usepackage{multicol, amsthm}
\usepackage{csquotes}
\usepackage{accents}
\newcommand{\ubar}[1]{\underaccent{\bar}{#1}}

\usepackage{parskip}
\usepackage{amsmath}
\usepackage{ulem}

\usepackage{epstopdf}
\usepackage[caption=false]{subfig}

\usepackage{setspace}
\usepackage[numbers,sort&compress]{natbib}
\bibpunct[, ]{[}{]}{,}{n}{,}{,}

\theoremstyle{plain}

\theoremstyle{definition}

\theoremstyle{remark}

\begin{document}

\normalem
\articletype{Original Research Paper}

\title{What to make of non-inferiority and equivalence testing with a \emph{post}-specified  margin?}

\author{
\name{Harlan Campbell\textsuperscript{a}\thanks{CONTACT Harlan Campbell. Email: harlan.campbell@stat.ubc.ca}, Paul Gustafson}
\affil{\textsuperscript{a}University of British Columbia Department of Statistics
Vancouver, BC, Canada, V6T 1Z2}
}

\maketitle

\begin{abstract}

In order to determine whether or not an effect is absent based on a statistical test, the recommended frequentist tool is the equivalence test.  Typically, it is expected that an appropriate equivalence margin has been specified before any data is observed.  Unfortunately, this can be a difficult task.  If the margin is too small, then the test's power will be substantially reduced.  If the margin is too large, any claims of equivalence or non-inferiority will be meaningless.  Moreover, it remains unclear how defining the margin afterwards will bias one's results.  In this short article, we consider a series of hypothetical scenarios in which the margin is  defined \emph{post}-hoc or is otherwise considered controversial.  We also review a number of relevant, potentially problematic actual studies from health research, with the aim of motivating a critical discussion as to what is acceptable and desirable in the reporting and interpretation of equivalence tests.

\end{abstract}




\pagebreak

\begin{singlespacing}
\small{\begin{quote}{\footnotesize{Facts do not accumulate on the blank slates of researchers' minds and data simply do not speak for themselves. [...] Interpretation can produce sound judgments or systematic error. Only hindsight will enable us to tell which has occurred.}}\end{quote}
\vspace{-0.5cm}
\hfill {\footnotesize{TJ Kaptchuk, 2003} \cite{kaptchuk2003effect}}}
\end{singlespacing}

\section{Introduction}

Consider the following hypothetical situation. After having collected data, we want to determine whether or not an effect is absent based on a statistical test.  All too often, in such a situation, non-significance (i.e. $p > 0.05$), or a combination of both non-significance and supposed  high power (i.e. a large sample size), is used as the basis for a claim that the effect is null. Unfortunately, such an argument is logically flawed.   As the saying goes, ``absence of evidence is not evidence of absence''~\cite{hartung1983absence, altman1995statistics}.  Instead, to correctly conclude the absence of an effect under the frequentist paradigm, the recommended tool is the equivalence test (also known as a ``non-inferiority test'' for one-sided testing \cite{wellek2010testing}), which tests whether an effect is at most constrained within a given equivalence margin.  However, it is generally accepted that we must specify the equivalence margin a-priori, i.e. before any data has been observed \cite{wellek2010testing}.  In our hypothetical situation, suppose that we did not have the foresight needed to have pre-specified this margin, are we then simply out of luck? 
 
It is worth noting that lack of foresight is only one reason we may have failed to have pre-specified an appropriate equivalence margin.  Defining and justifying the equivalence margin is one of the ``most difficult issues''~\cite{hung2005regulatory} for researchers.  If the margin we define is deemed too large, then any claim of equivalence will be considered meaningless. If the margin we define is somehow too small, then the probability of declaring equivalence will be substantially reduced~\cite{wiens2002choosing}.   While the margin is ideally chosen as a boundary to objectively exclude the smallest effect size of interest~\cite{lakensequivalence}, these ``ideal'' boundaries can be difficult to define, and there is generally no clear consensus among stakeholders~\cite{keefe2013defining}.  Furthermore, previously agreed-upon meaningful effect sizes may be difficult to ascertain as they are rarely specified in protocols and published results~\cite{djulbegovic2011optimism}. 

Suppose now that, having failed to pre-specify an adequate equivalence margin, we define the equivalence margin \emph{post}-hoc, having already collected and observed the data.  Given the potential consequences of interpreting data based on post-hoc decisions (e.g. see the ``Harkonen case'' as discussed in \cite{lee2016evaluating}), it is understandable that this idea may be alarming to some.  Hung et al. (2005)~\cite{hung2005regulatory} note that: ``If the margin can change depending on what has been observed [...] statistical testing of non-inferiority [or equivalence] may not be interpretable.''  And Wiens (2002)~\cite{wiens2002choosing} observes that: ``The potential biases of defining the margin after the study should be weighed against the cost and inconvenience of better understanding the differences [between study groups].''   Finally, the Committee for Proprietary Medicinal Products (CPMP) (the EU scientific advisory organization dealing with new human pharmaceuticals approval) \cite{committee2001points} notes that: ``it is prudent to specify a noninferiority margin in the protocol in order to avoid the serious difficulties that can arise from later selection.''

Statements such as these naturally lead one to ask the following: under what circumstances would equivalence testing with a data-dependent margin ``not be interpretable?''  What are the ``potential biases'' and ``serious difficulties''  we should consider in these, less than ideal, circumstances?  Walker (2011)~\cite{walker2011understanding} stresses that defining the equivalence margin before observing the data is ``essential to maintain the type I error at the desired level'' suggesting that potential type I error inflation is the issue of concern. Yet this too remains unclear.  This short article seeks to shed light on these curious questions by considering a series of rather confounding hypothetical scenarios (Sections 2 and 3) as well as a number of relevant, potentially problematic actual outcomes from health research, where equivalence and non-inferiority testing has been widely used for decades (Section 4).  We conclude (Section 5) with an invitation for further discussion about how best to address the title question: what to make of non-inferiority and equivalence testing with a \emph{post}-specified margin?

\section{The \emph{Pseudo}-Type I Error, the ``False Equivalence Rate'' (FER) and a pathological case}

Before going forward, we would be wise to recall that, under the frequentist paradigm, hypotheses are statements about parameters and therefore are nonrandom quantities. Hence, each hypothesis is either true or false, irrespective of how the data are realized.  

Let $\theta$ be the parameter of interest, let $X$ represent the data, and let $\ubar{\theta}(X;\alpha)$ and $\bar{\theta}(X;\alpha)$ be the lower and upper $(1-\alpha)\%$ confidence bounds, respectively, borrowing from the notation of \cite{wellek2017critical}.  Let us define a symmetric equivalence region as $(\theta_{0}-\Delta,\theta_{0}+\Delta)$, where $\theta_{0}$ represents the value of $\theta$ absent of any effect.   Without any loss of generality, let $\theta_{0}=0$. Then the standard equivalence testing hypotheses are defined as:

\noindent $H_{0}: \quad \theta \le - \Delta, \quad$ or $\quad \theta \ge  \Delta$ \\
$vs.$\\
$H_{1}: \quad -\Delta < \theta < \Delta$.

\begin{figure}\includegraphics[width=12.5cm]{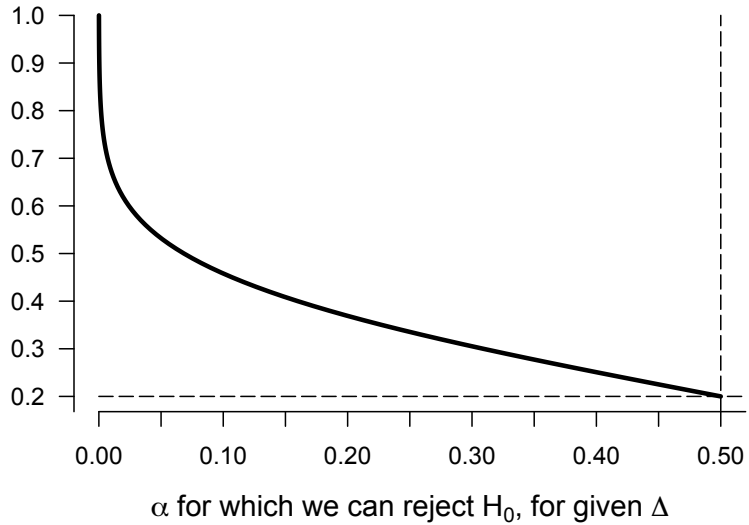} \caption{\textbf{The one-to-one correspondence between $\alpha$ and $\Delta$.}  In the above plot, an equivalence test is conducted on two sample normally distributed data.  The observed mean difference is $\hat{\theta} = 0.2$, and the observed pooled standard deviation is equal to 1.  The shape of this particular curve is specific to this particular data.  However,  for any general case, the smallest value of $\alpha$ needed to reject the null ($x$-axis) decreases as $\Delta$ increases ($y$-axis).  Furthermore, as the dashed lines indicate, when $\Delta=\hat{\theta}$, the corresponding value of $\alpha$ will be 0.5. }\label{fig:onetoone}
\end{figure}

\begin{figure}\includegraphics[width=12cm]{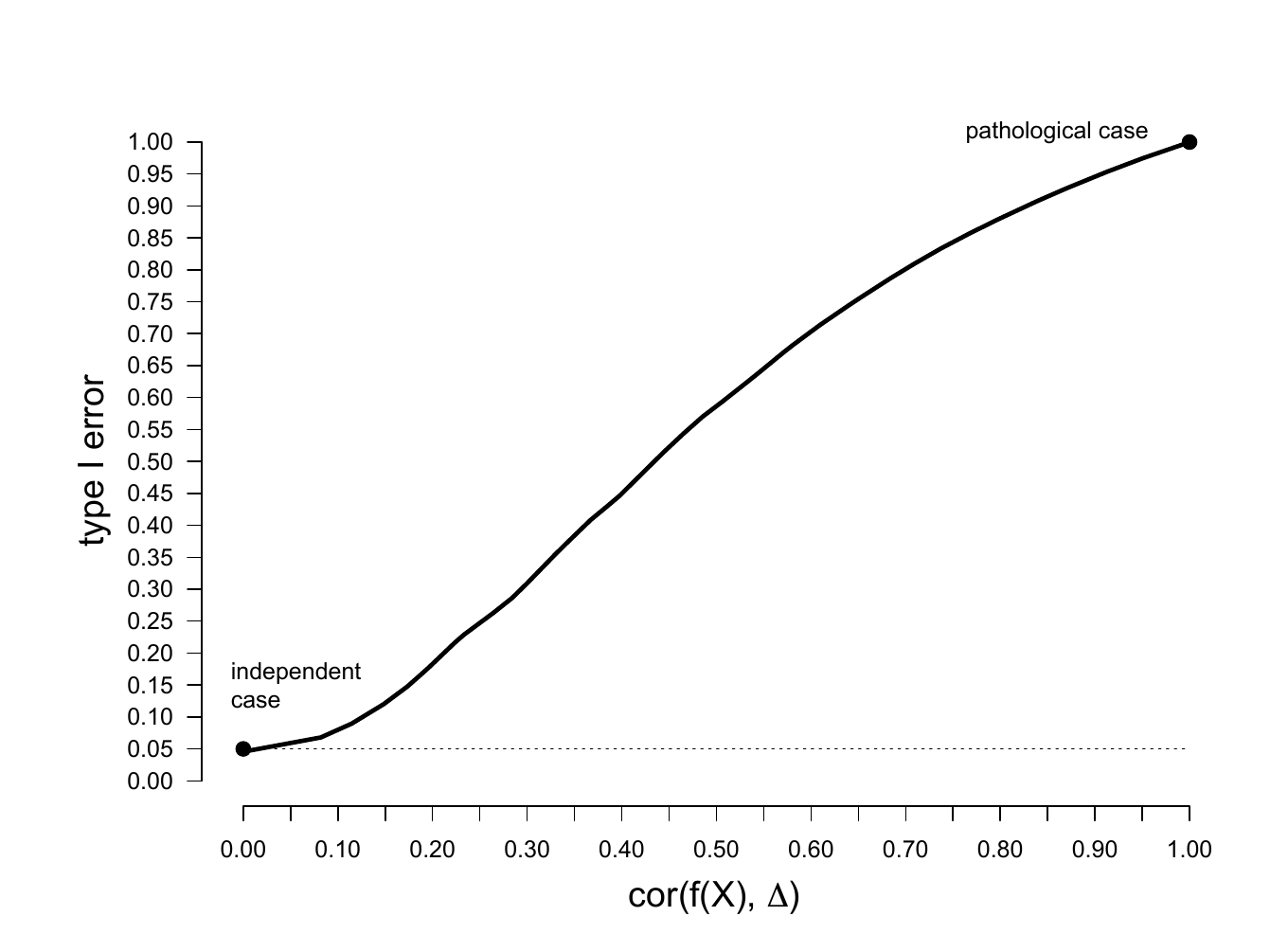}  \caption{\textbf{ In order for the test to be valid, the key is independence between the margin and the data.}  The relationship between type 1 error and the correlation between the margin and the data.  The curve is the result of repeated simulations of data, see details in Appendix.}
\end{figure}

 There is a one-to-one correspondence between symmetric confidence intervals and equivalence testing such that the null hypothesis, $H_{0}$, can be rejected whenever the realized confidence bounds satisfy $[\ubar{\theta}(X;2\alpha),\bar{\theta}(X;2\alpha)] \subset [-\Delta,\Delta]$.  Conversely, there will be insufficient evidence to reject the null hypothesis whenever $[\ubar{\theta}(X;2\alpha),\bar{\theta}(X;2\alpha)] \not\subset [-\Delta,\Delta]$.  Such a procedure is a valid test with type I error equal to $\alpha$.  In other words, we have the standard guarantee that $Pr(\textrm{reject }H_{0} | H_{0}\textrm{ is true}) \le \alpha$; see Westlake (1972) \cite{westlake1972use} and more recently Wellek (2017) \cite{wellek2017critical}.  

Should the equivalence margin not be specified a-priori, and be defined based on the observed data, we have the following admittedly \emph{improper} hypothesis test:

\noindent $\tilde{H}_{0}: \quad \theta \le - \Delta(X), \quad$ or $\quad \theta \ge  \Delta(X)$ \\
$vs.$\\
$\tilde{H}_{1}: \quad -\Delta(X) < \theta < \Delta(X)$.

\noindent In this case, we may not necessarily have that $Pr(\textrm{reject }\tilde{H}_{0} | \tilde{H}_{0} \textrm{ is true}) \le \alpha$.    To better understand, let us consider the following admittedly ``pathological case.'' Let  $\Delta(X)$ be chosen, based on the observed data, to be the smallest possible value for which one can claim equivalence (known in the literature as the ``LEAD'' boundaries, see Meyners (2007) \cite{meyners2007least}).  This is done by setting:

\noindent $\Delta(X) =  max(|\ubar{\theta}(X;2\alpha)|,|\bar{\theta}(X;2\alpha)|)+\epsilon, \quad$

\noindent  where $\epsilon$ is a small positive real number.
 
Note that, given the monotonic relationship between a confidence interval and an equivalence test, there is a one-to-one correspondence between $\alpha$ and $\Delta$. For any given value of $\alpha$, conditional on a fixed sample of data, there is a value for $\Delta$ for which one can reject $H_{0}$.  Conversely, for any given value of $\Delta$, there is a value of $\alpha$ for which one can reject $H_{0}$; see Figure \ref{fig:onetoone}.   
 
 In our pathological case, we have that $Pr(\textrm{reject }\tilde{H}_{0})=1$, i.e. we will always claim equivalence.      In this situation, the margin is entirely ``data-dependent.''  In other words, the data (as summarized by the confidence interval) and the margin are perfectly correlated.  We write $cor(f(X), \Delta)=1$, where $f(X)=max(|\ubar{\theta}(X;2\alpha)|,|\bar{\theta}(X;2\alpha)|)$.  Figure 2 displays the relationship between type 1 error and $cor(f(X), \Delta)$, see details in the Appendix.   In the pathological case, since $Pr(\textrm{reject }\tilde{H}_{0})=1$, we also have that $Pr(\textrm{reject }\tilde{H}_{0} | \tilde{H}_{0}) = 1$.  As such, we have $Pr(\textrm{reject }\tilde{H}_{0} | \tilde{H}_{0}) > \alpha$, and therefore, the ``\emph{pseudo-}type I error'' is not controlled.  When there is less correlation, i.e. when the margin is not entirely data-dependent, we can expect to see less type 1 error inflation.  In order for the test to be valid, the key is independence between the margin and the data.  In the case when the data and the margin are entirely independent, the type 1 error rate will be equal to $\alpha$, as desired.

 Oddly enough, the probability that the ``improper null hypothesis'' is true, $Pr(\tilde{H}_{0})$, is at most, equal to $\alpha$.  One can easily understand why this is problematic from a frequentist point of view.  As stated earlier, under the frequentist paradigm, a hypothesis is either true or false with probability 0 or 1.   The only way to conceptualize the ``probability of the null hypothesis'' is to adopt a \emph{pseudo}-Bayesian philosophy.  That being said, let us momentarily suspend any such qualms, and consider $Pr(\tilde{H}_{0} | \textrm{reject }\tilde{H}_{0})$.  Following from the definition of the confidence interval, we have that:

$Pr(\tilde{H}_{0} \mid \textrm{reject }\tilde{H}_{0}) \le \alpha$.

\noindent Proof:
\vspace{-1.3cm}
%
%
%
%

\begin{align*} 
Pr(\tilde{H}_{0} \mid \textrm{reject }\tilde{H}_{0}) &= Pr(\tilde{H}_{0}) & \textrm{  (because we will always reject $H_{0}$)}\\
&= Pr (|\theta| > \Delta(X))&\\
&= Pr(|\theta| >  max(|\ubar{\theta}(X;2\alpha)|,|\bar{\theta}(X;2\alpha)|)+\epsilon) &\textrm{(by definition of $H_{0}$)}\\
&\le \alpha &\textrm{ (by definition of a confidence interval)}
\end{align*}

If the concept  of ``$Pr(\tilde{H}_{0} | \textrm{reject }\tilde{H}_{0})$'' seems familiar, it may be because this resembles the false discovery rate (FDR) often considered when conducting multiple comparisons.  Since, in this particular case, a ``discovery'' is in fact a claim of equivalence, let us instead refer to this rate as the ``false equivalence rate'' (FER).

In summary, for this worst-case-scenario ``pathological case'', while the ``\emph{pseudo}-type 1 error rate'' is not controlled (and is in fact equal to 1), we can guarantee that FER is at most equal to $\alpha$.  Whether or not this is useful for a given claim of equivalence or non-inferiority depends entirely on the practical (e.g. clinical) implications.  Is knowledge that, with a controlled FER, the magnitude of the effect is at most equal to the margin $\Delta(X)$ sufficient?  Equivalently, is knowledge that the confidence interval includes the null value and is sufficiently narrow adequate?   

Just as in standard equivalence testing (i.e. when the margin is pre-specified), if the margin is too generous, then any claim of equivalence (regardless of type I error or FER) will be practically meaningless.  The question is therefore: what constitutes ``too generous''?  Or what is ``sufficiently narrow''?  As we shall see in Section 4, researchers, public opinion, subject matter experts, policy makers, and regulatory agencies may all have their own opinions.

\section{A somewhat less pathological case}

Let us briefly consider a somewhat less pathological case.  The CPMP published an advisory report, ``Points to consider on switching between superiority and non-inferiority'' \cite{committee2001points},  in which they describe another hypothetical situation where the margin is determined after the data is observed: 

\begin{displayquote}
``Let us suppose that a bioequivalence trial finds a 90\% confidence interval for the relative bioavailability of a new formulation that ranges from 0.90 to 1.15. Can we only conclude that the relative bioavailability lies between the conventional limits of 0.80 and 1.25 because these were the predefined equivalence margins? Or can we conclude that it lies between 0.90 and 1.15?

The narrower interval based on the actual data is the appropriate one to accept. Hence, if the regulatory requirement changed to +/- 15\%, this study would have produced satisfactory results. There is no question here of a data-derived selection process.

However, if the trial had resulted in a confidence interval ranging from 0.75 to 1.20, then a post hoc change of equivalence margins to +/-25\% would not be acceptable because of the obvious conclusion that the equivalence margin was chosen to fit the data.''
\end{displayquote}

According to this recommendation, it seems that, without any scrutiny, we are free to shrink a pre-specified margin as needed.  However, we should always avoid widening the pre-specified margin if that is what is necessary. If this is the case, it would suggest that a prudent strategy would be to always pre-specify the largest possible margin before collecting data, and then shrink the margin as required.  This may strike some as opportunistic and potentially problematic.

Ng (2003) \cite{ng2003issues} studies a similar ``somewhat less pathological'' case in which a large, possibly infinite number of margins are all pre-specified and all the corresponding hypotheses are tested (without any Bonferroni-type of adjustment for multiple comparisons).   Equivalence is then claimed using the narrowest of all potential pre-specified margins for which equivalence is statistically significant.  Ng (2003) \cite{ng2003issues} explains why this hypothetical strategy may be problematic: ``Although there is no inflation of the type I error rate [due to the fact that all hypotheses are nested], simultaneous testing of many nested null hypotheses is problematic in a confirmatory trial because the probability of confirming the finding of such testing in a second trial would approach 0.5 as the number of nested null hypotheses approaches infinity.''  

To better understand Ng (2003) \cite{ng2003issues}'s concern, consider a similar setup where, for a standard null hypothesis significance test, the $\alpha$-level (allowable type I error) is data-dependent, always set to be equal to the value of the observed $p$-value $+ $ $\epsilon$, where $\epsilon$ is some small constant.  Then, the probability of confirming the statistical significant finding in a second trial (with identical sample size and $\alpha$) approaches 0.5 as $\epsilon$ approaches 0; see Hoenig and Heisey (2001) \cite{hoenig2001abuse}.  As such, it is always expected that one specifies (and justifies) the chosen $\alpha$-level prior to observing any data; see the recent related commentary of \cite{lakens2018justify}. (These two situations are in fact identical, due to the aforementioned one-to-one correspondence between a data-driven selection of $\alpha$ and a data-driven choice of $\Delta$; see Figure \ref{fig:onetoone}.)

\section{How hypothetical are situations like these?}

While the cases described in the previous sections were purely hypothetical, similar situations do arise in practice.  We list a number of different studies as examples, with the aim of motivating a critical discussion as to what is acceptable and desirable in the reporting and interpretation of equivalence tests.

First, consider cases of post-hoc judgement that often arise in the regulatory approval of drugs seeking a designation of bio-equivalence for approval.  When the pre-specified margin is deemed too generous (i.e. too wide) by regulatory authorities only after the data have already been observed and analyzed, the regulator may decide that for the purposes of approval, the drug does not meet an appropriate standard for equivalence.  Consider two examples:

\begin{enumerate}

\item{ The {SPORTIF III and SPORTIF V} randomized controlled trials (RCTs) were studies designed to investigate the potential of ximelagatran as the first oral alternative to warfarin in patients with nonvalvular atrial fibrillation to reduce the risk of thromboembolic complications.  The primary end point in each study was the incidence of all strokes and systemic embolic events, and the primary objective was to establish the non-inferiority of ximelagatran relative to warfarin with a pre-specified margin of an absolute 2\% difference in the event rate; see \cite{halperin2003ximelagatran}.

Both studies met the primary objectives of non-inferiority with the pre-specified margin. As such, upon completion, the studies were heralded as a ``major breakthrough'' \cite{kulbertus2003sportif, albers2005ximelagatran}.  However, upon regulatory review by the FDA Cardiovascular and Renal Drug Advisory Committee, the pre-specified margin was judged to be ``too generous'' \cite{boudes2006challenges}.  This post-hoc criticism of the ``unreasonably generous'' \cite{kaul2005trials} margin, along with concerns about potential liver toxicity, led the FDA to refuse to grant approval of ximelagatran for any of the proposed indications, see Head (2012) \cite{head2012non}.


}

\item{ The {EVEREST II study} was a RCT designed to evaluate percutaneous mitral valve repair relative to mitral valve surgery \cite{mauri2010everest}.  The primary efficacy end point was defined as the proportion of patients free from death, surgery for valve dysfunction, and with moderate-severe (3+) or severe (4+) mitral regurgitation at 12 months.  
Upon completion, researchers claimed success when  the primary non-inferiority objective was achieved.  However, the conclusion of non-inferiority was ``difficult to accept due to unduly wide margins''  \cite{head2012non}.  Thus, the FDA determined that despite the highly significant $p$-value, ``non-inferiority is not implied due to the large margin'' and therefore the data ``did not demonstrate an appropriate benefit-risk profile when compared to standard mitral valve surgery and were inadequate to support approval'' \cite{fda2013}. }

\end{enumerate}

In other instances, the complete opposite occurs.  Despite the fact that the researchers fail to pre-specify a specific margin prior to observing the data, the regulatory agency may still accept a claim of equivalence/non-inferiority on the basis that, given some non-controversial post-hoc margin, there is sufficient evidence.  Consider two examples:

\begin{enumerate}
\item{ The goal of MannKind's {``Study 103''} was to evaluate the inhaled insulin Afrezza for the treatment of diabetes mellitus in adults.  Subjects were randomized to 12 weeks of continued treatment in one of three treatment arms.   The pre-specified primary objective was to show superiority of the Afrezza TI+metformin arm relative to the secretagogue+metformin arm, with respect to change in HbA1c at 12 weeks.  Upon completion, the superiority objective was not achieved and a non-inferiority margin had not been pre-specified by the researchers.  However, the regulators were able to accept a claim of non-inferiority.  The FDA clinical review states (Yanoff, 2014) \cite{yanoff2014}:  ``The sponsor did not specify a non-inferiority margin. However, the FDA statistical reviewer noted that Afrezza TI+metformin was non-inferior to secretagogue+metformin when the standard margin of 0.4\% for insulins is used (the upper bound of the 95\% confidence interval for the treatment difference in HbA1c is 0.3\%).''
 }
 
 \item{{The  {ALLY-3 trial} was a one-arm phase 3 trial with the goal of evaluating the safety and efficacy of oral daclatasvir for chronic HCV genotype 3 infection \cite{mccormack2015daclatasvir}.  There was no active or placebo control and as such it was impossible to conduct a non-inferiority or equivalence test based only on the trial data.  As such the FDA looked to other trials to determine estimates for the effectiveness of competitor treatments.   In addition, as noted by the Oregon Health Authority \cite{herink2012class}, ``[t]he ALLY-3 trial [...] did not define a non-inferiority margin for determination of efficacy. The FDA analysis calculated it based on historical data and concluded that DCV [daclatasvir] with SOF [sofosbuvir] achieved non-inferiority compared to SOF [sofosbuvir] with RBV [ribavirin] for 24 weeks[...].'' In this case, the FDA reviewers ``clinically justified'' their choice of a \emph{post}-specified non-inferiority margin based on a historical data; see \cite{struble2015}.}
}

\end{enumerate}

\noindent These studies illustrates the fact that, in some fields, there may be well-established ``standard'' margins or sufficient ``historical data.'' Such standards no doubt make \emph{post}-specification less controversial. 

The appropriateness of the margin is also often debated in peer-reviewed journals, both before and after publication (e.g. the post-hoc debate between Groenewoud et al. \cite{groenewoud2017response} and Gupta et al.  \cite{gupta2016impact} about the appropriateness of the pre-specified non-inferiority margin defined in Groenewoud et al. (2016) \cite{groenewoud2016randomized}'s study on methods for embryo transfer). Consider two cases of interest.   In the first case, the margin was not pre-defined, yet claims of equivalence were nevertheless put forward.  In the second case, while a margin was pre-defined, additional conclusions were made based on \emph{post}-specified margins.

 \begin{enumerate}
 
\item{ {Chang et al. (2008)} \cite{chang20085} published the results of a RCT with the goal of evaluating a 5- versus 3-day course of oral corticosteroids (CS) for children with asthma exacerbations who are not hospitalised.  The primary outcome was 2-week morbidity of children.  The study did not show a statistically significant difference between the two treatment arms. In the interpretation of the results, Chang et al. (2008) note that: ``It would have been ideal to define a non-inferiority or equivalence margin a priori on the basis of a minimally important effect or historical controls. Our study was designed as a superiority trial, and we did not define a non-inferiority margin a priori. Nevertheless, for the primary outcome measure, the chosen symptom score cut-off of 0.20 (i.e., chosen minimally important difference), the study shows equivalence (the upper 95\% confidence limit of the difference between groups in our study was 0.18).''  As such, the researchers concluded that the 3-day and 5-day treatment courses were ``equally efficacious'' in reducing the symptoms of asthma \cite{chang2007longer}.}

\item{ {Jones (2016)} studied the efficacy of isoflurane relative to sevoflurane in cardiac surgery \cite{jones2016comparison}.  When interpreting the results, the authors note that: ``our choice of non-inferiority margin may seem to be overly generous; however, it is important to emphasize that, if the margin had been reduced to as low as 1.5\%, the conclusions of this trial would not have changed.'' \cite{jones2016comparison}.}

\end{enumerate}

\section{Conclusions}

In clinical trials research, expectations that a margin be pre-specified have been well established for quite some time \cite{piaggio2006reporting}.  This is not the case in other disciplines.  In psychology and the behavioural, educational, and social sciences, where the practice of equivalence testing is ``rapidly expanding'' \cite{koh2013robust}, discussions of how best to execute equivalence tests are underway and appropriate recommendations are crucially needed.  Indeed, in many disciplines, researchers now advocate that equivalence testing has great potential to ``facilitate theory falsification'' \cite{quintana2018revisiting}.  By clearly distinguishing between what is ``evidence of absence'' versus what is an ``absence of evidence'', equivalence testing may facilitate the long ``series of searching questions'' necessary to evaluate a ``failed outcome'' \cite{pocock2016primary}.  As a result, it may encourage greater publication of null results which is desperately needed \cite{fanelli2011negative}.  Yet, outside of health research, guidelines on how best to define and interpret margins are lacking.  We hope that the title question of this article will motivate researchers to further consider the delicate issues involved. 

One might argue that the pathological case of equivalence testing does not actually qualify as \emph{testing} per se, and is instead, simply a tool for \emph{describing} the data.  This is the opinion of Meyners (2007), who concludes that, as a descriptor of the data, the ``LEAD boundaries'',  $[-\Delta(X),\Delta(X)]$, provide ``useful information'' and in some cases are ``even more important than confidence intervals'' for reporting results \cite{meyners2007least}.

At the end of the day, everyone must arrive at their own conclusions as to whether or not a sufficient standard of evidence for equivalence has been demonstrated.  Obviously this is often easier said than done.  An infamous example from the clinical trials field is the debate over using bevacizumab (avastin) as a treatment for age-related macular degeneration.  A non-inferiority study was conducted to investigate \cite{catt2011ranibizumab}.  However, some considered the pre-specified non-inferiority margin of 5 letters  as ``generous'' even before the results of the trial were announced \cite{reuters2011}.  This suggests that, regardless of the results, some would have remained skeptical of any claim of non-inferiority with the 5-letter margin.  In stark contrast, the standard of evidence for others was much weaker: many doctors determined that the use of bevacizumab (avastin) as a substitute for ranibizumab (lucentis) was justified (particularly given the ``too big to ignore'' price difference) even before the completion of the non-inferiority trial and were comfortable treating large numbers of patients with Avastin ``off-label'' \cite{steinbrook2006price}.  In this situation, financial incentives clearly played a competing role with clinical efficacy in what was to be considered ``equivalent.''

While the use of equivalence testing should be encouraged, caution is warranted.  In a review of equivalence and non-inferiority trials, Le Henanff et al. (2006) \cite{le2006quality} find that often studies ``reported margins [that] were so large that they were clearly unconvincing.''  Indeed, as G{\o}tzsche et al. (2006) conclude: ``clinicians should especially bear in mind that noninferiority margins are often far too large to be clinically meaningful and that a claim of equivalence may also be misleading if a trial has not been conducted to an appropriately high standard.'' We conclude with the following general recommendations:

\begin{itemize}

\item The validity of an equivalence test does not depend on the margin being pre-specified.  Rather, the necessary requirement for a valid test is that the margin is completely independent of the data.  Furthermore, simply because a margin has been pre-specified (and is therefore guaranteed to be independent of the data), it is not necessarily an appropriate choice.  Regardless of whether the margin is pre-specified, or defined \emph{post}-hoc, we must acknowledge that a claim of ``noninferiority [or equivalence] is almost certain with lenient noninferiority margins'' \cite{flacco2016noninferiority}.  

\item {If one is to suggest equivalence based on a \emph{post-hoc} margin, one must, at the very least, be forthcoming and honest about the potential for bias.  In such cases, every effort should be made to justify the appropriateness of the \emph{post}-specified margin based on factors entirely independent of the observed data and to acknowledge the potential for outside criticism.}

\item {In the absence of a pre-specified margin, one can always resort to simply reporting the associated confidence interval. If the confidence interval contains the null and is ``narrow enough,'' the absence of an effect can be deemed likely. This tactic lacks the formalism of equivalence testing, yet avoids the difficulties with the interpretation and justification of a post-hoc margin.}

\item {Researchers, given their incentive to publish \cite{nosek2012scientific}, are not in the best position to define their own margins.  This is true whenever the margin is pre-specified, and especially true when a margin is suggested post-hoc.  As such, in order to avoid any potential scrutiny, researchers would be wise to seek an independent party, void of any potential biases, to define an appropriate margin.  This is already common practice in clinical trial research, where sponsors have undeniable incentives to further drug development and the FDA and other regulators will (ideally) set a guidance and layout expectations for an acceptable margin.  In other fields, the suggestion that an equivalence margin be defined/scrutinized by an independent party has recently been considered within the framework of a proposed publication policy.  In the conditional equivalence testing publication policy, the independent journal editor/reviewers are tasked with critically evaluating a given margin prior to the start of a study \cite{campbell2018conditional}.}

\end{itemize}

\vskip 0.52in

\bibliographystyle{tfs}
\bibliography{truthinscience} 

\section{Appendix}

Details of Figure 2.  The plotted curve is the result of repeated simulations of two-sample normally distributed data.  The details of the simulation are as follows.

We generate 50,000 simulations for each unique value of $p$, as selected from an equally spaced sequence ranging from 0 to 1. For each simulation, we proceed through the following five steps:

\begin{enumerate}
\item{Two independent samples ($n=50$) of data are generated from $Normal(0,1)$ and $Normal(\mu,1)$ distributions respectively.}

\item{A two-sided 90\% confidence interval, $[\ubar{\theta}(X;0.10), \bar{\theta}(X;0.10)]$, is calculated for the difference in population means.}
	
\item{The binary variable $\pi$ is generated from a $Bernoulli(p)$ random variable such that, $\pi=1$ with probability $p$, and  $\pi=0$ with probability $1-p$.}

 \item{If $\pi=0$, $\Delta$ is randomly generated from a $HalfNormal(\mu-\epsilon,0.01)$ distribution so that its value is somewhat random but always less than $\mu$.}
 
 \item{If $\pi=1$,  $\Delta$ is set to equal to $max(|\ubar{\theta}(X;0.10)|,|\bar{\theta}(X;0.10)|)$, as in the ``pathological case.''}
 \end{enumerate}
 
The quantity $cor(f(X), \Delta)$ is based on the observed correlation as calculated from all simulations for each given value of $p$.  Naturally, larger values of $p$ correspond to higher degrees of correlation. For the plotted curve, we set $\mu=0.5$, and $\epsilon=0.001$.

\end{document}